# Development of a new model to predict indoor daylighting : integration in CODYRUN software and validation

A. H. Fakra, F. Miranville, H. Boyer, S. Guichard

Physics and Mathematical Engineering Laboratory for Energy and Environment (P.I.M.E.N.T)
University of La Reunion
117 rue du Général Ailleret, 97430 Le Tampon (Reunion Island – French overseas Dpt)
fakra@univ-reunion.fr

**Abstract**

Many models exist in the scientific literature for determining indoor daylighting values. They are classified in three categories: numerical, simplified and empirical models. Nevertheless, each of these categories of models are not convenient for every application. Indeed, the numerical model requires high calculation time; conditions of use of the simplified models are limited, and experimental models need not only important financial resources but also a perfect control of experimental devices (e.g. scale model), as well as climatic characteristics of the location (e.g. in situ experiment).

In this article, a new model based on a combination of multiple simplified models is established. The objective is to improve this category of model. The originality of our paper relies on the coupling of several simplified models of indoor daylighting calculations. The accuracy of the simulation code, introduced into CODYRUN software to simulate correctly indoor illuminance, is then verified.

Besides, the software consists of a numerical building simulation code, developed in the Physics and Mathematical Engineering Laboratory for Energy and Environment (P.I.M.E.N.T) at the University of Reunion.

Initially dedicated to the thermal, airflow and hydrous phenomena in the buildings, the software has been completed for the calculation of indoor daylighting. New models and algorithms - which rely on a semi-detailed approach - will be presented in this paper.

In order to validate the accuracy of the integrated models, many test cases have been considered as analytical, inter-software comparisons and experimental comparisons. In order to prove the accuracy of the new model – which can properly simulate the illuminance – a confrontation between the results obtained from the software (developed in this research paper) and the major made at a given place is described in details. A new statistical indicator to appreciate the margins of errors - named RSD (Reliability of Software Degrees) – is also be defined.

The objective is not only to develop an efficient research tool to improve visual comfort and reduce energy consumption, but also to transfer the knowledge through these decision-making aids tools to praticians and decision makers.

**Keywords**: Indoor daylighting, building simulation software, experimental validation, CODYRUN.

**Contents:**

**Nomenclature**

| | |
|---|---|
| DF | Daylight Factor (%) |
| ERC | Externally Reflected Component (unitless) |
| $E_{in,dif,p}$ | Indoor global Diffuse illuminance at point p (lux) |
| $E_{in,Dir,p}$ | Indoor global Direct illuminance at point p (lux) |
| $E_{mean}$ | Illuminance Mean value of the dependent variable testing data set and N (number of records of data in the testing set) or the $E_{meas,i}$ mean values (lux) |
| $E_{meas,i}$ | Measured value for horizontal illuminance (lux) |
| $E_{mod,i}$ | Illuminance predicted values (lux) |
| $E_{out,Dir,p}$ | Direct illuminance from the sun at point p (lux) |
| $E_{out,Dir,S}$ | Direct illuminance from the sun at point p (lux) |
| $E_{out,G,p}$ | Outdoor Diffuse illuminance at point p (lux) |
| FC | Correction caused by the remoteness of a point illuminated by natural light from an opening (unitless) |
| FR | Correction factor for window framing (unitless) |
| GL | Daylight transmission coefficient of the glass (unitless) |
| IRC | Internally Reflected Component (unitless) |
| MF | Correction factor for window dirt (unitless) |
| MG | Activity coefficient of the study site (unitless) |
| SC | Sky Component (unitless) |
| $S_{TS}$ | Surface of the sun patch (m$^2$) |
| $S_T$ | Floor surface (m$^2$) |
| $V_{sim,i}$ | Interior simulated illuminance value at a point of position index i (lux) |
| $V_{ref,i}$ | Interior reference illuminance value at a point of position index i (lux) |
| $R^2$ | Coefficient of correlation |

*Greek Symbol*

| | |
|---|---|
| $\rho_{sol}$ | Indoor average slab reflectance (unitless) |
| $\tau_{vitre}$ | Coefficient of light transmission of glazing (unitless) |

*Index*

| | |
|---|---|
| i | Positioning index of a point on a horizontal plan surface indoor the building. i varies from 0 to N. |

## 1. Introduction

CODYRUN software is a dynamic building simulation code developed at the Physics and Mathematical Engineering Laboratory for Energy and Environment (P.I.M.E.N.T). The software uses a multiple-model approach to simulate the energy behaviour of buildings, including thermal, aeraulic and hydrous aspects. It was in its first release created in 1993 [1] for the needs of two types of users:

- *Building designers and operators*
- *Building physics researchers*

Thereafter, several works have been investigated to improve the computer tool. Many physical models were tested and introduced into the software to make it more powerful and more diversified.

Recently, a new model has been developed and integrated to allow the simulation of indoor daylighting. The objective is to simulate illuminance at any point of a work plane by using a new concept to simplify the combination of photometric models. This model is original because it is based on the combination of three simplified types of models to determinate indoor illuminance.

It is indeed important to provide a tool capable of simulating daylighting, simultaneously coupled with thermal phenomena. This allows studying cases involving interactions between several energetic phenomena and determining the global behaviour of buildings.

## 2. Literature review

The literature review will be only conducted on the quantity of indoor daylighting coming from sidelights toward a point located on a specified useful work plane.

### 2.1. History

The CIE (International Commission of Illumination) was created in 1928 [2] in order to answer to the problem associated with daylighting and lighting. The objective of the organization is to develop norms, measurement methods and models to characterize luminous environment [3]. It is recognized by International Standardisation Organisms due to its know-how and professionalism in illuminance analysis applications.

From 1960, Kittler [4] introduced artificial skies with the aim of simulating indoor daylighting of scale models. Thus, more complex indoor daylighting configurations could be studied under normalised experimental conditions.

In 1973 [5], the CIE published a new methodology to assess indoor daylighting from Daylight Factor (DF). Some fifty other methods are presented in this same report.

Tregenza and Waters [6] introduced the notion of Daylight Coefficient (DC) in 1983.

Collins [7] marked the history and discoveries in the evaluation of indoor daylighting. In fact, in his article, published in 1984, he reminded us that the very first research on quantitative study of daylighting was present since 1865. The studies were made at that time from a graphical approach to be used with simple rectangular sidelight (without considering complex cases).

In the same year, numerical calculation codes in daylighting and lighting have progressed in computers technology, more precisely, with the development of computers and software: Superlite [8], Radiance [9], Genelux [10], Lightscape 3.2 [11], Adeline [12], etc.

In 1988, Clarke and Janak [13] improved RADIANCE and ESP-r to determine indoor daylight from Daylight Factor (or DF).

In 1999, Michel [14] introduced a new factor for the calculation of indoor daylighting, called the Partial DF.

In the same year, Paule [15] created a simulation code (DIAL) for study of indoor daylighting and lighting. The characteristic of this software is that it allowed preliminary draft diagnosis of a building using simplified and accurate models (minimal calculation time).

In 2004, Maamari [16] defined a series of test cases to verify the reliability of numerical simulation software applicable to indoor natural and artificial lighting.

Currently, many laboratories and organizations across the world are working to improve numerical simulation tools and modelling dedicated to indoor illuminance. This is the case, for example, of the models developed by Enrique Ruiz in 2002 [17], Jenkins and al. in 2005 [18], and Danny in 2007 [19], or the task 31 of IEA in 2007 [20].

2.2. State of the art – Daylighting, choice and selected models

Characterising indoor daylighting is very difficult because various parameters must be considered (view factor, aperture size, depth of the local, etc.).

Generally, there are three types of methods to determine quantitative indoor illuminance:

- *Experimental (scale models, full-sized buildings, etc.);*
- *Numerical (Radiosity, Ray tracing, etc.);*
- *Simplified (Lumen, DF, etc.).*

Experimental models require important financial means (scanning sky simulator, for example, in the case of studies of daylighting from scaled models) and the perfect observation of experimental phenomena, through measurements, on scale models or full-sized buildings. The second drawback in this method is that the time delay to obtain exploitable data is relatively long and depends on the climatic constraints of the site. Concerning scale models, errors are very high when compared to results obtained by full-scale measurements. This approach is therefore to be avoided.

In the case of numerical detailed models, calculation time is considerable. In addition, these models are generally used for visual rendering (quality of the illuminance) rather than for quantitative characteristics of daylight in a room.

Finally, the major inconvenience of simplified models is that they are only applicable under certain conditions (sky conditions, limited to a certain position, etc.). However, they have enormous advantages: easy to understand and to implement, simple to use and less time consuming, give acceptable results, etc. These models are more adaptable to the code source of CODYRUN. Therefore, this category has been selected for integration of a daylighting model in CODYRUN. Our contribution in the field has been to combine three types of models normally used in different conditions (overcast sky or not, only direct sunlight, etc.). This allowed not only to circumvent the conditions calculations, but also to improve their applications range.

### 3. Modelling of indoor daylighting

In this paragraph, new models and algorithms for the numerical simulation of daylighting in CODYRUN are presented.

3.1. Brief description of CODYRUN software

CODYRUN is a multi-zone software integrating natural ventilation, thermal and moisture transfers, developed with a friendly interface.

The software is a tool dedicated to researchers and professionals. Physical models taken into account in the software are:

- ✓ Airflow and humidity transfer
- ✓ Outdoor convection
- ✓ Indoor convection
- ✓ Outdoor long wave exchanges
- ✓ Indoor long wave exchanges

- ✓ Indoor short wave repartition
- ✓ Heat conduction
- ✓ Conductive exchanges with the ground
- ✓ HVAC system
- ✓ Outdoor illuminance
- ✓ $CO_2$ transport

Details of the multiple-model characteristics of CODYRUN are given in [21], the thermal model constitution in [22], the pressure airflow model in [23], the outdoor illuminance model in [40], the data-structuration and the description of the tool in [24] and $CO_2$ model in [25].

Many validation tests of CODYRUN code were successfully applied to the software. Most of them are part of the BESTEST procedure [26] and led to the validation of the results.

3.2. Implementation of a new model

3.2.1. Introduction

All these models allow the software to address not only the various issues related to building design but also to make a detailed study of the physical variables of the surrounding.

However, the initial model does not study other aspects related to the interior environment, such as visual comfort. An improvement of the software was necessary to overcome this.

Studies have thus been conducted to integrate indoor daylighting models in the software and ensure that the software correctly simulates illuminance received at any point of a given work plane (see Figure 1).

This study is original because many research laboratories are only interested in the qualitative study of the relationship between thermal transfer and luminous aspects (for example, the association of a temperature for each luminous colour) as opposed to CODYRUN, which is able to examine quantitative values.

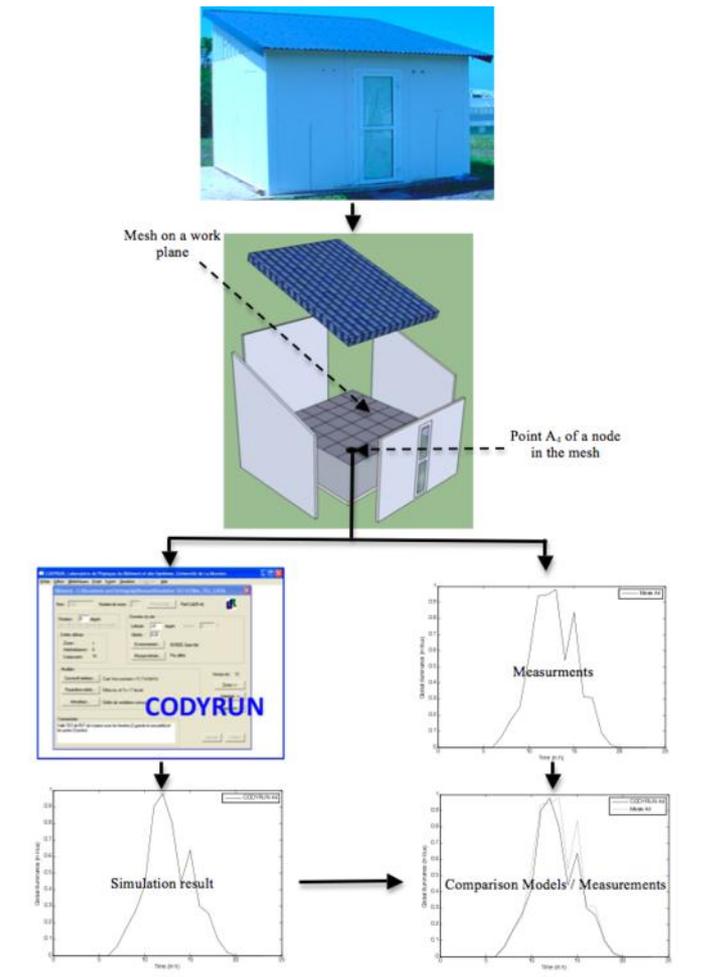

**Figure 1. Objective of the study**

3.2.2. Improvement of the software

Many improvements have been made to the software CODYRUN to enable it to properly simulate indoor daylighting. Figure 2 summarizes the various computer codes that have been developed in this study. The objective is to develop a 3-D toolkit for building environments dedicated to the daylighting simulation. The tools to represent this environment are algorithms and data structures in space vector (vector, polygon and projection). These algorithms are often used in programming of any sort of 3D computer graphics.

For this, the computer code (program) of CODYRUN has been amended at four different levels:

- ✓ In the simulation code: it was necessary to add indoor daylighting model, to create a work plane and to make the mesh of the plane from an algorithm based on Euclidean geometry (polygons, vertex, vector, etc.) and finally integrate a model capable of reproducing solar illuminance from a meteorological database containing only irradiance (global, diffuse and direct beam).

- ✓ <u>In the result files</u>: CODYRUN creates an Excel file where it usually stores all the information on the results of simulations. This file is not suitable in the photometric study. So, two improvements have been made at this level: first, the existing result file has been modified to take into account the value of illumination on the working plane. On the other hand, new file storage information, much more detailed in the field of lighting, was added. This file (output variable of the software) can be opened with a notepad application.

- ✓ <u>In the weather file</u>: the numerical simulation of a building cannot be done without prior knowledge of the weather conditions. Indeed, CODYRUN needs to be informed about climatic conditions (input variables of the software). Regarding the simulation of the interior illuminance, the software needs to have information on the illumination normally provided by external data format files such as TMY2. CODYRUN accepts such files. Sometimes, we do not have a similar database. Knowing that the software has its own type of weather file but does not give any information on illuminance variable, an improvement has therefore been made to provide information on outdoor illuminance. In the lack of weather data, it is possible to simulate indoor illuminance. Indeed, at a given time it is sufficient to indicate an overall value of outdoor illuminance for the calculations of the indoor illuminance at any point on a work plane.

- ✓ <u>In the building description</u>: another input variable should be given to the software: the description of the studied building. However, the indoor illuminance strongly depends on the geometric position in space of openings, building orientation, etc. Previously, the software could not take into account all these parameters. This problem was solved by adding the cartesian coordinates in the declaration of each component (wall, floor, ceiling, windows, etc.) of the building. Each component is constituted of two sides and each side is characterized by one vertex.

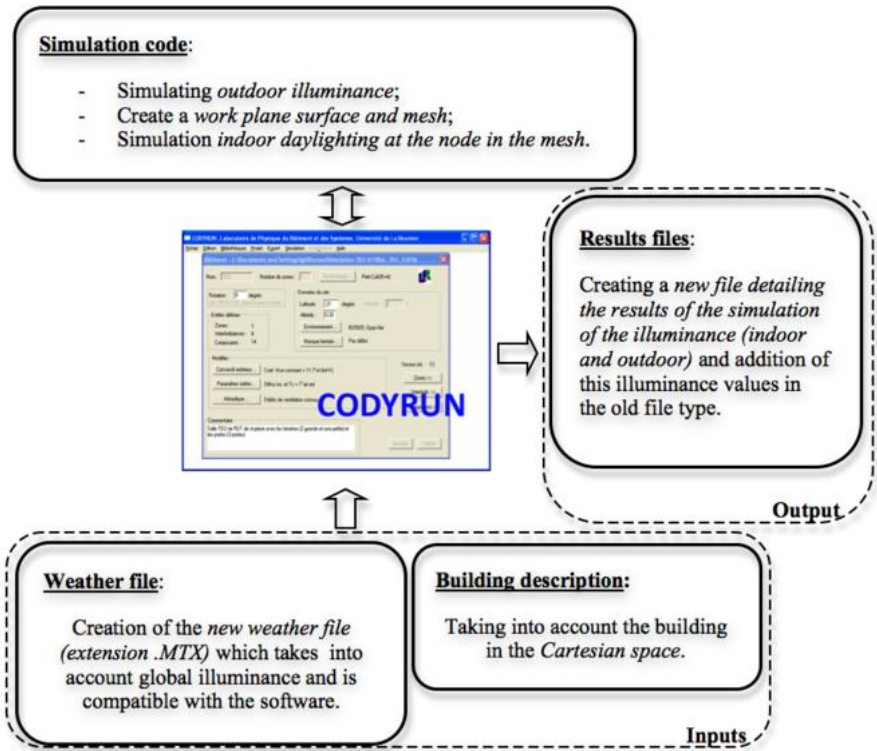

**Figure 2. Improvements to the software CODYRUN**

3.2.3. Description of the model

Two basic models are used in the algorithmic combination of the calculation code for indoor daylighting. The Daylight Factor (or DF) is used to determine diffused illuminance whereas the classical method (taking into account point position and aperture visible transmittance) is used to determine the direct illuminance through a sidelight. The principle of calculation for each one of these two methods will be given in the following paragraphs. In addition, a new specific mathematical relation for the determination of the diffuse illuminance on a sun patch will be described.

*3.2.3.1. Calculation of diffuse illuminance in the absence of sun patch: DF method*

The DF (Daylight Factor) method will be used to calculate diffuse illuminance. This method was elaborated by BRE (Building Research Establishment) [27] and published by CIBSE [28]. We can consider the light falling on a point in the rooms as being composed of three distinct components. Light that comes directly from the sky and called the Sky Component (SC), light that comes from external surfaces such as buildings called the Externally Reflected Component (ERC), and light that is reflected from internal surfaces called the Internally Reflected Component (IRC). The sky component normally refers to the diffuse sky: it is not used to describe direct sunlight. The DF is simply calculated as:

$$DF = (SC + ERC + IRC \times FC) \times MF \times FR \times GL \times MG \qquad (1)$$

If a DF value (from the relation (1)) is known, it is possible to calculate the global inside illuminance at the point of interest from the relation (2):

$$E_{in,dif,p} = DF \times E_{out,G,p} \qquad (2)$$

*3.2.3.2. Calculation of diffuse illuminance part in the presence of the sun patch : new method*

It is considered that the diffuse illuminance at a point (which is inside the sun patch) is the sum of the diffuse illuminance generated by the DF and that produced by the sun patch itself in the neighbourhood of the point. A simple relation originally determines the diffuse part coming from the sun patch:

$$E_{in,dif,p} = E_{out,Dir,S} \times \rho_{sol} \times (S_{TS} / S_T) \qquad (3)$$

The general equation of diffuse illuminance at an inside point (of the room) on the sun patch becomes:

$$E_{in,dif,p} = (DF \times E_{out,G,p}) + (E_{out,Dir,S} \times \rho_{sol} \times (S_{TS} / S_T)) \qquad (4)$$

*3.2.3.3. Calculation of the sunlight*

For this part, the Siret's method is used [29]. The relation is written as followed:

$$E_{in,Dir,p} = E_{Dir,S} \times \tau_{vitre} \qquad (5)$$

*3.2.3.4. Recapitulation: new method of indoor illuminance calculation*

Relations (2), (3) and (5) allow establishing a new combination model of simplified categories to calculate indoor illuminance for every condition:

$$E_{glo,p} = (DF \times E_{out,G,p}) + (E_{out,Dir,S} \times \rho_{sol} \times (S_{TS} / S_T)) + E_{Dir,S} \times \tau_{vitre} \qquad (6)$$

It is important to note that this simple new relation is original since (2), (3) and (5) are usually used separately (for each according to different conditions of application). Whereas the new relation found (6) is the combination of the three models. Thanks to this new combination, it is possible to calculate the illuminance values for any sky conditions. Indeed, (2) is applied only in overcast sky condition, while (3) and (5) are applied if and only if a light flux - coming from the sun - enters the room. The scientific literature does not give information on the combination of relations (2), (3) and (5), which are combined together (i.e. relation (6)).

## 4. Validation

The validation step is very important to know the precision of the simulation results. Especially when dealing with daylighting simulation programs, the achievement of the accuracy of the predictions is difficult. It is thus necessary to refer to rigorous procedures, recognized worldwide.

There are few documents on the procedures to follow. In most cases, laboratories implement their own experimental database to serve as a reference for comparisons between model predictions and measurements.

The study of Maamari [30] perfectly responds to our needs concerning the validation step. Indeed, his thesis is related to the establishment of methods to check the reliability of simulation codes of indoor daylighting and artificial lighting, from analytical and experimental tests cases. This work is used as reference in the task TC3-33 of the CIE [31].

Many other test cases found in the scientific literature (papers from CSTB, BRE, Task 21 of IEA and experimental test case of CIBSE, etc.) have been applied to the simulation software CODYRUN. An inter-software comparison has also been done and finally, database references for the local study in dynamic conditions (from a cell called LGI) have been established. This last case will only be presented here.

### 4.1. Preliminary

#### 4.1.1. Limitations of models used in CODYRUN for indoor daylighting simulation

CODYRUN only provides daylight values for a given horizontal surface (i.e. no daylighting values are calculated on vertical walls).

The software only considers sidelight. In addition, it does not take into account internal obstructions (furnishing, occupants, etc.), bidirectional transmissions (BRDF) and lightwell. Therefore, simulation results will be confronted to reference test cases, taking into account the above-said hypothesis.

#### 4.1.2. New approach to determine the percentage of reliability

It is important to consider a square percentage deviation to calculate the error between measurements and used models, when studying outdoor or/and indoor daylighting.

##### *4.1.2.1. Outdoor errors*

We have recently taken into account used models for the study of luminous efficacy [32]. It concerns the references [33] and [34].

Generally, three statically indicators are considered to characterise the accuracy of the models associated to outdoor and indoor daylight: the coefficient of determination $R^2$, the Mean Bias Deviation (MBD) and the Root

Mean Square Deviation (RMSD). MBD demonstrates the model's tendency to underestimation or to overestimation. RMSD offers a deviation measure from the predicted values in relation to the measured values. The use of the correlation coefficients $R^2$, MBD, and RMSD determines the accuracy of the models. The following relations give the definitions of these statistical estimators:

$$RMSD = \left(\frac{1}{E_{mean}}\right) \times \left[\frac{\sum_{i=1}^{N}\left(E_{mod,i} - E_{meas,i}\right)^2}{N}\right] \quad (6)$$

$$MBD = \left[\frac{\sum_{i=1}^{N}\left(E_{mod,i} - E_{meas,i}\right)}{N \times E_{mean}}\right] \times 100 \quad (7)$$

In order to increase accuracy, some statistical indicators also need to be defined as follows:

$$R^2 = \left[\frac{\sum_{i=1}^{N}\left(E_{mod,i} - E_{meas,i}\right)^2}{\sum_{i=1}^{N}\left(E_{meas,i} - E_{mean,i}\right)^2}\right] \quad (8)$$

*4.1.2.2. Indoor errors*

Concerning indoor daylighting study, the estimate is made from relative errors ε at the point i defined by:

$$MBD = \frac{V_{sim,i} - V_{ref,i}}{\left|V_{ref,i}\right|} \quad (9)$$

Global relative (ε) errors are thus written:

$$\varepsilon = \frac{1}{N} \times \sum_{i=1}^{N} \varepsilon_i \quad (10)$$

*4.1.2.3. Reliability of the Software Degrees (RSD): new concept of quantification of the accuracy*

In this study, a new percentage indicator that verifies the capacity of the software to correctly simulate indoor daylighting is defined as RSD (or Reliability of Software Degree). It is the ability of the software to simulate correctly and realistically a physical variable. In our case study, the variable is indoor daylighting.

Photometric phenomena are very complex, so we can consider that results of numerical simulation software are correct compared to reality (or reference values) when its RSD is greater or equal to 50%. This gives a precise idea of the limits, strengths and weaknesses of the software in its application field. As a preliminary, this method requires the knowledge of reference values (or reference test case study).

Two situations are presented for the determination of RSD:

- On the one hand, the reference values are given at each measurement. For example, positioning reference point $A_1$ at $A_4$ (or reference calculation) in form of margin (upper and lower limits of the curve below) of acceptable values for the results (results given by CODYRUN on the curve below) of the simulation. These margins are given by measured values more or less the total estimated error with and without errors linked to the simulation (see figure 3). In this case, the RSD is the percentage of simulated values situated between references values.

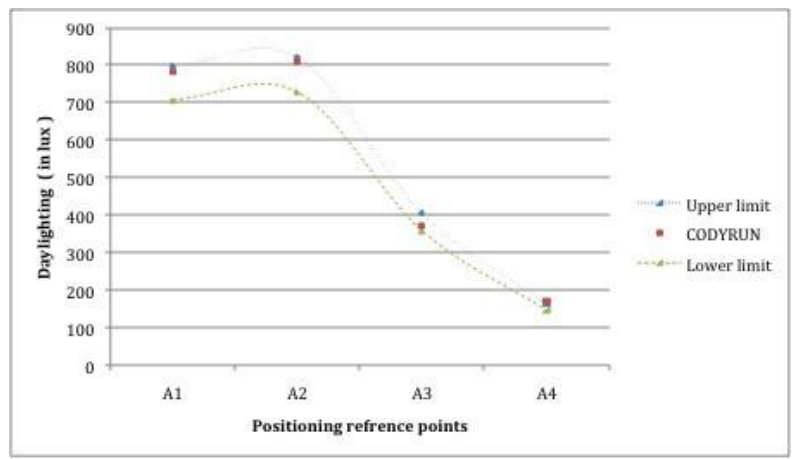

**Figure 3. Illustration of RSD evaluation when reference points are margins (upper and lower limits)**

- On the other hand, reference values are given at each measurement point (or reference calculation) without margins. So, it is essential to consider relative mean errors $\varepsilon$ between simulation and reference values at each point. The RSD is given by subtracting 100% from the global mean relative error obtained (refer to figure 4).

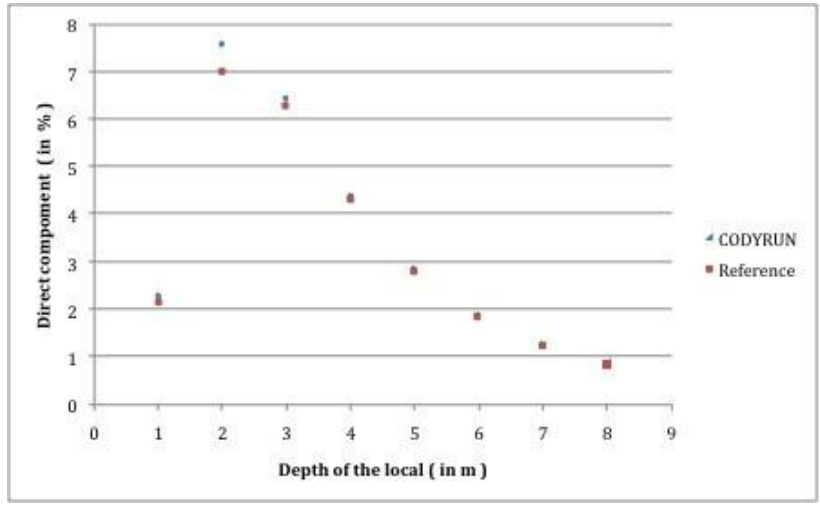

**Figure 4. Illustration of RSD evaluation when reference points are given without margins**

4.1.3.  Tests cases selection of the C.I.E

The reference scenarios of the C.I.E are classified into two categories: analytical and experimental. In each category, we can distinguish the study of artificial lighting and/or daylighting. These references are defined in the Task TC3-33 of the CIE. They are used for the validation of CODYRUN. Only 39% of these tests cases were applied to CODYRUN and verified.

4.2. Summary and results of test cases submitted to CODYRUN

Table 1 summarises all test cases applicable to CODYRUN and associated errors for each case. CODYRUN successfully passed selected test cases (the errors were acceptable).

Globally, results for CODYRUN are comparable to those of other codes for daylighting. This shows that the model included in CODYRUN is well implemented and gives acceptable results.

Results of analytical test cases confirm the aptitude and the limitations of the software to simulate daylighting. Unlike the approximate values of results concerning the calculation analytical test case, the simulation of CODYRUN code gives exact values. Consequently, this difference of calculation increases the errors in the comparison between the simulation in CODYRUN and the analytical test case.

Concerning the references of experimental test cases, it is important to know that they are very complicated to implement, requiring a lot of logistics, technical and financial means. For example, accuracy for spatial location of the sensors is very important; it is also important to ensure that the measureable limits of the sensors corresponds to the range of measured values.

As a conclusion, this work allowed us to assess the strengths and weaknesses of the software. Thus, the improvement of the software could be done on the illuminance calculation at the furthest and nearest point to the apertures.

| Test Type | Reference | Procedure | Relative error (%) | RSD (%) | Observation |
|---|---|---|---|---|---|
| Analytical | [30] | II.10 | 3.87 | 91.44 | Side opening |
| | [30] | II.11 | 4.54 | 93.65 | Different types of skies CIE |
| | [30] | II.14.4 | 1.96 | 98.04 | Diffuse (form factor) |
| | [30] | II.14.9 | 15.61 | 84.34 | Solar irradiance (outdoor) |
| Comparative | [37] | CSTB | 5.92 | 95 | Daylight Factor (DF) |
| Experimental | [30] | III.2.5 | 3.38 | 93.49 | Artificial sky |
| | [36] | I.E.A | 9.31 | 74.46 | Scale model E.P.F.L. |
| | [38] | I.E.A | 11.31 | 70.92 | Real Building H.U.T. |
| | [35] | LGI test | 32 | 88.95 | Transient (or dynamical aspect) |
| | | Minimum = | 1.96 | 84.34 | |
| | | Maximum = | 32 | 98.04 | |
| | | Average = | 9.76 | 87.81 | |

**Table 1. Reference test cases applied to CODYRUN**

4.3. Example of dynamic validation: experimentation on LGI cell

Table 1 shows the details of test cases submitted to the software CODYRUN to check its level of reliability to simulate indoor daylighting.

The highest error between the reference values and those from the simulations were observed during the dynamic study performed locally using the LGI cell (about 32%) of relative error [35]. The following paragraph will present a detailed study to explain the results.

4.3.1. Experimental test cell LGI

An existing experimental full-scale test cell has been instrumented in order to verify the reliability of the software CODYRUN to correctly simulate daylighting in transient conditions during an entire day. The database collected is used as a reference for the comparison between simulation and experimental values.

The experimental test cell LGI is located at Saint-Pierre (see figure 5). This test building was built for experimental validation reference on physical models introduced in CODYRUN [10] and ISOLAB [39].

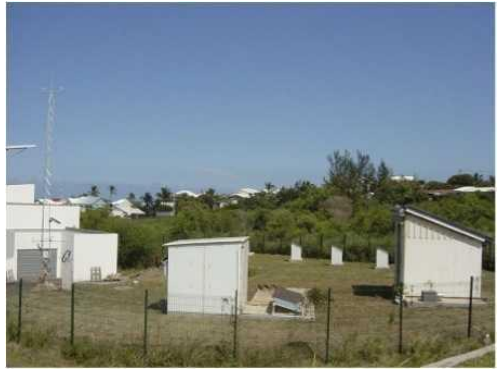

**Figure 5. Experimental site for L.P.B.S. Laboratory**

Two types of luxmeters were used for the global illuminance: one to characterize the outdoor daylighting and another to characterize the penetration of indoor daylight.

The experimental platform includes a meteorological station and a technical room. The internal and external walls were initially painted white. Absorption coefficient of the interior surfaces is about 60 %. The roof is constituted of corrugated iron and insulated with a radiant barrier. The vertical walls are made of insulating board. The floor is composed of concrete slabs, of polystyrene on the top (4 and 5 cm thick). The dimensions of the LGI cell are illustrated in the figure 8.

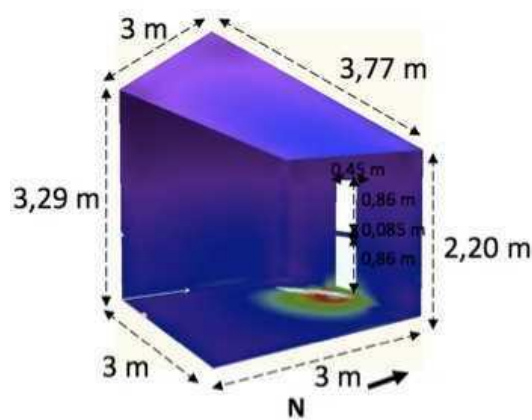

**Figure 6. LGI cell test of L.P.B.S.**

The LGI cell is representative of a typical room or office in Reunion Island. The door is made up of aluminium, with a 6 mm glass window.

4.3.2. Instrumentation and experimental procedure

The meteorological data are measured every minute. The station measures irradiation (global and diffuse) and wind speed and direction (respectively at a height of 2 m and 10 m from the ground). We added an outdoor daylight sensor that measures more than 14,000 values each day. The data logger is of CAMPBELL type. The measurements of indoor and outdoor daylighting are made from instruments of AHLBORN type. All the measuring instruments are synchronised. To simplify our study, we have positioned an exterior luxmeter (sensor) on the roof of the LGI cell (see figure 8). Indoors luxmeters (sensors) were aligned on the slabs as shown in the figure (see figure 9). These sensors are equidistant by 0.5 m. The horizontal distance between position of the sensor $A_1$ and aperture is 0.23 m.

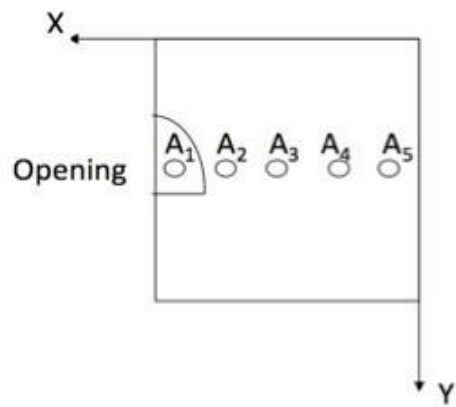

**Figure 7. Schematic position of the five interior luxmeters**

Globally, three categories of days (refer to [40] for more information) are used as a reference experiment: it consists of intermediate day, clear day and overcast day conditions. The reference Table 2 gives the subjective and objective measurements errors. Total error associated with the experimental measurements (outdoor and indoor measurements) is 13.22% and the global error is 15%.

4.3.3. Measurement errors

Table 2 indicates the values of subjective and objective errors of our experiment. This approach to study the measurement error is highly recommended by the C.I.E [31].

These values will allow us to give margins of lower and upper limits for each measurement. The mean error (measurements errors) is 13.4% and global error (measurements and software hypothesis of simulation errors) is 15%. Table 3 gives the characteristics of the interior measurements sensors.

| Error | Identifications | Error value (en %) |
|---|---|---|
| Objective | Sensor accuracy | 1 (indoor) + 0.03 (datalogger) + 9.9 (outdoor) |
| | Cosine correction | 2.9 (indoor sensor) + 2,9 (outdoor sensors) |
| | Spectral sensitivity | 1 (indoor sensor) + 7 (outdoor sensor) |
| Subjective | Accuracy on positions and orientation of the measuring points | 3 |

**Table 2. Subjective and objective errors of experimentation of the LGI cell test**

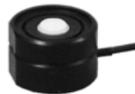

| Measuring range | 0.05 – 12 500 lux |
|---|---|
| Spectral sensitivity | Class B, superior at 6% |
| Cosine error | < 3 % |
| Rated temperature | 24°C +/- 2 K |
| Range of humidity | 10 to 90% (without condensation) |
| Operating/storage temperature | 0°C to +60°C / -10°C to +80°C |

**Table 3. Characteristics of the interior measurements sensors (FLA603VL2), of AHLBORN type**

4.3.4. Simulation hypothesis in CODYRUN

The simulation conditions of the LGI cell in CODYRUN are:

- Albedo = 70%

- Interior reflectance of the walls and ceiling (white colour) = 60%

- Interior reflectance of the floor (grey-coloured slabs) = 20%

- The height from the slabs to the workplane is at 0.01 m.

- Dimension of the elementary surface grid = 0.1m x 0.1m

- Correction factor of DF: MG = FR = 0.8 and MF = 0.9 (see (1) for the definition of these coefficients)

4.3.5. Comparison of results

This paragraph presents the simulation results and the comparison with those of measured indoor daylighting obtained for the three days conditions.

- **Clear day condition**

Figure 8 indicates the comparison between measured values and simulated values at point $A_3$. The trend at points $A_1$, $A_2$, $A_4$, and $A_5$ are practically the same, but with different amplitudes.

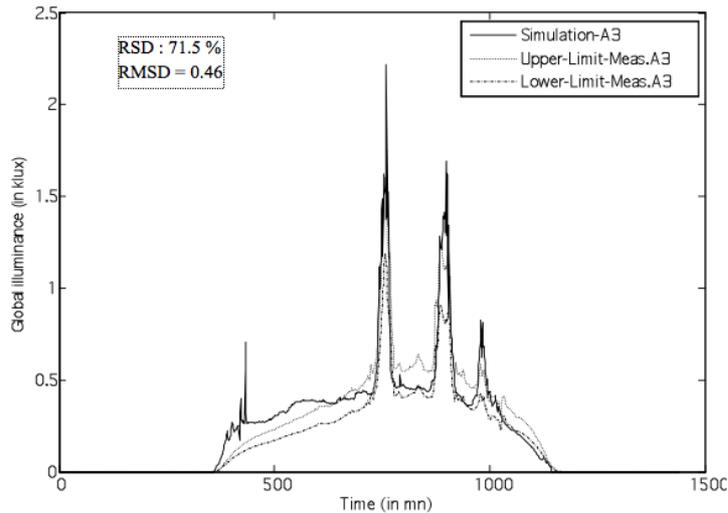

**Figure 8. Clear day: comparison between measured and simulated values at the point $A_3$ (without direct sunlight and for one-minute time step)**

We obtained a RSD value of more than 50%. The simulation results obtained - without considering the direct sunlight flux - better correspond to the values of measurements (see Figure 8). There is obviously no inertia in photometric phenomena (i.e. abrupt variations of daylighting with time). So, during our measurements we took a time step of one minute to take into consideration these abrupt variations. The inconvenience is that this increases the errors between the simulation results and reference measurements. To verify this hypothesis, we plotted the same curve with hourly averaged values and we got the following figure (refer to Figure 9, for the same point $A_3$). Reference values obtained hourly present more realistic results than minutely simulated results.

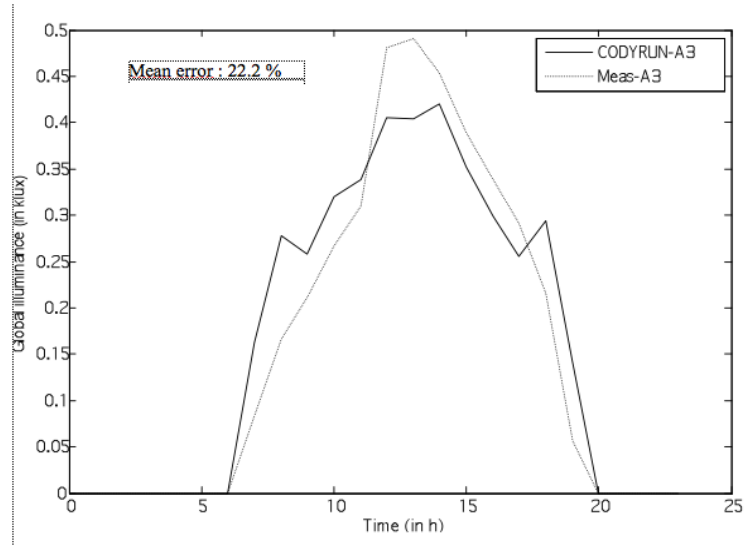

**Figure 9. Clear day: comparison between measured and simulated values at the point A₃ (for one-hour time step)**

- **Intermediate day**

The comparison results of position of the reference point at $A_1$ are given next. Figure 10 shows an example of comparison between simulations and measured values. The results obtained for other positions are quite identical and the observation is the same as for clear day. CODYRUN has a RSD equal to 71.3 %.

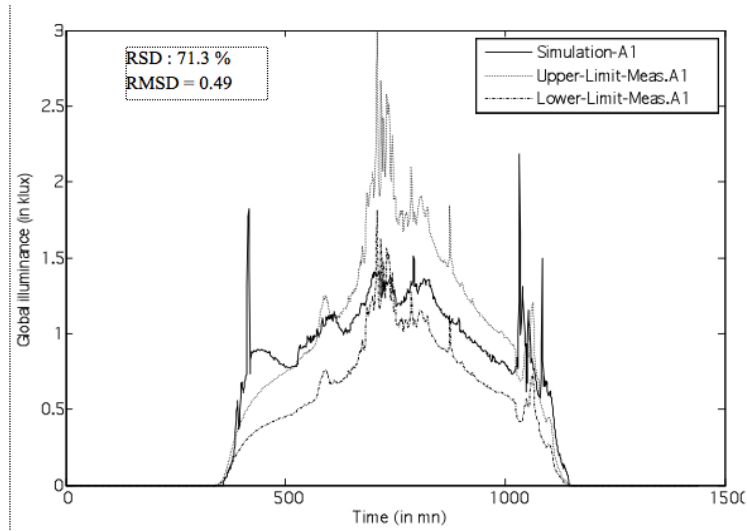

**Figure 10. Intermediate day: comparison between measured and simulated values at the point A₁ (for one-minute time step)**

Figure 11 compares simulation results with results for an hourly time step at point $A_1$. That day was divided into two equal parts: a relatively clear morning and a cloudy evening. During some precise moments of the morning, very high illuminance measurements have been observed, due to direct sun penetration. Indeed, the interior sensors were not able to measure the direct illuminance flux of values which were superior to 12 500 lux because of their measurements range, which does not go beyond this value (See Table 3). This explains why the software seems to simulate incorrectly the global available illuminance in the morning. In fact, simulation results in the evening better correspond to actual values.

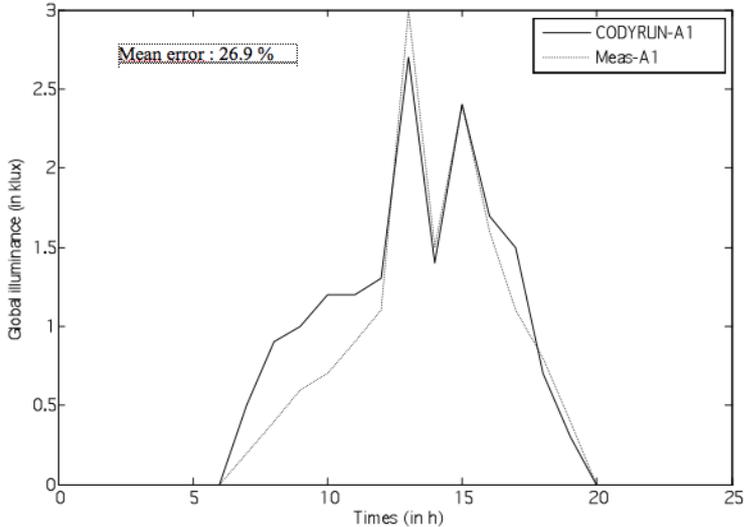

**Figure 11. Intermediate day: comparison between measured and simulated values at the point $A_1$ (one-hour time step)**

- **Overcast day**

In this study, we present simulation results obtained from reference positions points $A_5$. Figure 12 illustrates the results of simulation.

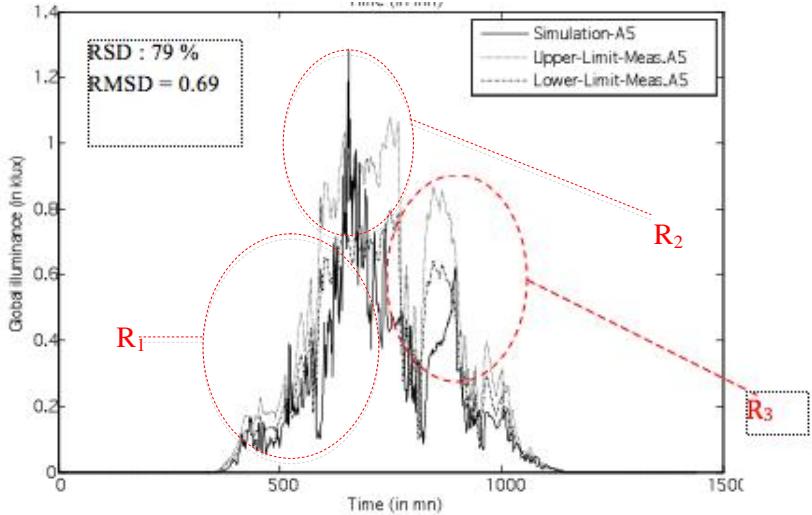

**Figure 12. Indoor daylighting : Overcast day condition (LGI cell)**

In most cases, simulated values in CODYRUN between 10:30 am and 12:45 pm are higher compared to measured values (for example see the highlighted zone $R_2$ in the Figure 12). The explanation of this result is that the studied cell is equipped with an overhang on the northern façade. This prevents a major part of the sunlight from penetrating into the room. At this time, this particular device is not taken into account by the software. We noted that (for the five reference positions) a significant difference between simulation and measurement values exists in the morning (highlighted region $R_1$) compared to the evening (highlighted region $R_3$):

- ✓ In fact, from sunrise till 12:00 am (corresponding to highlighted $R_1$ zone in the Figure 12) CODYRUN better simulates the indoor daylighting. This is certainly due to the recognition of obstacles by the software (buildings in the neighbourhood of the north-east façade corresponding to the sun rising façade). This has been taken into account for indoor daylighting by the software;

- ✓ Concerning the evening, we noticed that (for example, the highlighted $R_3$ zone in $A_5$ - Figure 12) the simulated values are lower than the measured values. The fact is that South and West parts of vertical façades of the test building are opaque. Then, the North-facing overhang does not influence skylight at these hours of the day. We obtain more coherent values between the simulation and the measurements.

As previously, we were interested in comparing these reference values with those simulated values for each hour. The results (see the result comparison of figure 13 at reference point $A_1$) show a good accuracy of the software in the calculation of indoor daylighting, with the exception of close and far positions from the opening.

To conclude, in the case of an overcast day we can note that reliability accuracy is higher or equal to 68%. Globally, CODYRUN has the tendency to underestimate indoor daylighting.

4.3.6. Conclusion of the transient study

The case of a clear day study demonstrated the difficulty of simulating transient indoor daylighting. Indeed, the photometric effect is very complex (it is difficult to try to draw the shape of the illuminance for a given day and at a given point). For this reason, many other research laboratories prefer to conduct a time step study, but not during an entire day (see [25] and [26]). However, we found that the simplified models introduced in the software were able to simulate fairly well the dynamic photometric variable. Similar situations were presented in an intermediate day study.

For an overcast day, we noted that the impact of the overhang in the simulation was important. Indeed, the model introduced in CODYRUN does not consider, for the moment, the influence of overhangs. The fact that the overhangs are not considered in the simulation increases the errors of comparisons in the experimental validation. We also noted that the software made higher errors at reference points located close and far from the opening.

During an entire day, the RMSD of the overall simulation is about 0.61. Considering the relative errors of theses hourly simulations, we obtain an average error of 32.7%.

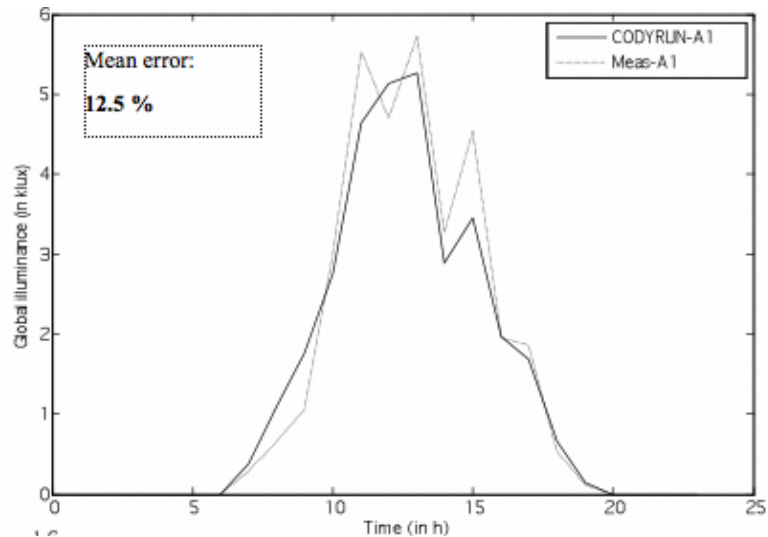

**Figure 13. Overcast day: hourly comparison between measured and simulated values at the points A$_1$**

## 5. Conclusions

The sun is a renewable energy source often used in interior lighting (inside buildings).

This allows a decrease in the electrical energy consumption for artificial lighting in buildings. But, we do not have enough information for the study concerning the needs of illuminance in tropical and humid climate regions.

In the building, the integration of daylighting coming directly or indirectly from the sun is a variable, which is more and more used in development tools to determine the illuminance availability. However, these tools must satisfy the needs of architects and building design offices. According to them, the tool should not only quantify daylight, but should also elaborate method tools, which are easily usable. CODYRUN tries to satisfy this requirement.

As for any other software, the use of CODYRUN to simulate indoor daylighting presents advantages and drawbacks.

Concerning the advantages, we can indicate:

- ✓ Reduced calculation time. For example, the software takes 2 minutes to simulate daylighting on a workplane of 39x35 position points inside a building and for a time step, it takes one minute all long year);

- The software is simple, user-friendly, and is perfectly adapted to design office (engineering consulting firm) and to individuals;
- This software is a tool used in research laboratories. We were able to show that CODYRUN was reliable at 86.25%;
- It is possible to reconstitute the outdoor illuminance from a meteorological file, which only has irradiance information;
- It is possible to simulate the effects of sunlight coming from a secondary light source;
- We can monitor (or follow) the sun patch in the workplane;
- CODYRUN can separately simulate the direct, diffuse and global daylighting for every sky condition, through a new simplified model that is defined in this paper;
- The software is capable of dynamical studies.

The limitations of the model introduced in the software are:
- It does not consider overhangs and shadow masks in daylight simulation;
- No daylight calculation on vertical or inclined walls are considered;
- The cartesian coordinates of the building are manually inserted. This makes the description of the building (in CODYRUN) a time-consuming task;
- The are no graphical user interface, and no 3D visual representation of the building or visual rendering;
- The used models are simplified categories, so they are not applicable to complex architectural configurations;
- It does not consider obstructions inside the room (furniture, occupants, etc.) when simulating daylighting;
- The opening should only be sidelights.

The conditions, which can be used in the software, are those of simple (the volumes are rectangular, triangular, cubic, or L-shaped) closed systems (or sub-systems). We cannot simulate complex systems (for example, inclined walls of buildings).

## 6. Future research

The future research in this field can be classified in three categories as follows:
- The integration of an indoor artificial lighting model and the improvement of the new indoor daylighting model in CODYRUN tool;

- ✓ Complementary validation results, e.g. performed once again the experimental study for intermediate and clear condition days in LGI test cell by using sensors capable of taking into account indoor direct sunlight;
- ✓ Using the tool to optimise the electrical energy consumption and to consider models taking into account simultaneously the temperature and the daylight.

**Vitae**

Dr Ali Hamada FAKRA

Laboratoire Physique et Ingénierie Mathématique pour l'Energie et l'environneNT (P.I.M.E.N.T)

Faculté de Sciences de l'Homme et de l'Environnement

Université de La Réunion - Campus Sud

17, Rue du Général Ailleret - 97430 Tampon - Ile de la Réunion – France

Email: fakra@univ-reunion.fr

Tel : +262 692 56 23 95          Fax : +262 262 57 94 46